\documentclass[%
  reprint,
  superscriptaddress,
  amsmath,amssymb,
  aps,
  floatfix
]{revtex4-2}

\usepackage{graphicx}
\usepackage{physics}
\usepackage[colorlinks=true]{hyperref} 
\usepackage{tabularx}
 
\draft
\begin{document}

\title{Radial etching of strongly confined crystal-phase defined quantum dots}

\author{Markus Aspegren}
\email[]{markus.aspegren@ftf.lth.se}
\affiliation{Division of Solid State Physics and NanoLund, Department of Physics, Lund University, P.O. Box 118, SE-221 00 Lund, Sweden}

\author{Chris Mkolongo}
\affiliation{Division of Solid State Physics and NanoLund, Department of Physics, Lund University, P.O. Box 118, SE-221 00 Lund, Sweden}

\author{Sebastian Lehmann}
\affiliation{Division of Solid State Physics and NanoLund, Department of Physics, Lund University, P.O. Box 118, SE-221 00 Lund, Sweden}

\author{Kimberly Dick}
\affiliation{Centre for Analysis and Synthesis and NanoLund, Department of Chemistry, Lund University, P.O. Box 124, SE-221 00 Lund, Sweden}

\author{Adam Burke}
\affiliation{Division of Solid State Physics and NanoLund, Department of Physics, Lund University, P.O. Box 118, SE-221 00 Lund, Sweden}

\author{Claes Thelander}
\affiliation{Division of Solid State Physics and NanoLund, Department of Physics, Lund University, P.O. Box 118, SE-221 00 Lund, Sweden}

\date{\today}

\begin{abstract}
We realize strongly confined quantum dots (QDs) in InAs nanowires (NWs) by combining epitaxial crystal-phase control with chemical wet etching. 
    A strong axial confinement is first introduced by growing closely spaced wurtzite (WZ) tunnel barriers in NWs to enclose a zinc blende (ZB) QD. 
    The NW cross-section is then reduced by isotropic etching to obtain very small QDs, with a maximum observed charging energy $> 30$ meV.
Using low-temperature electrical characterization and finite-element method simulations, we study how charging energies and the onset of electron filling scale with QD diameter.
    For extremely small diameters, we identify a regime where stray capacitances become non-negligible, limiting further increase in charging energy by diameter reduction alone. 
This approach to increasing confinement is particularly relevant for understanding the strong spin-orbit interaction observed in crystal-phase QDs, possibly linked to polarization charges at the WZ/ZB interfaces. 
Small diameter QDs allow considerably weaker interfering electric fields when studied, but the QDs cannot be realized with epitaxial growth alone due to a loss of crystal phase control.

\end{abstract}

\maketitle

\section{Introduction}
Quantum dots (QDs) confine charge carriers to nanometer dimensions, discretizing the energy spectrum and enabling controlled studies of few-electron physics \cite{Hanson2007}.
They have been realized in a wide range of material systems and geometries, such as lithographically defined QDs in two-dimensional electron gases \cite{Tarucha1996, Fujisawa2002, Petta2005}, self-assembled and colloidal nanocrystals \cite{Murray1993, Bimberg1999}, carbon-based systems \cite{Guttinger2012, Laird2015}, and quasi-one-dimensional semiconductor nanowires (NWs) \cite{Fasth2005}.

Semiconducting NWs are a particularly attractive starting point for QD applications, as additional confinement only needs to be introduced in the axial direction, thereby providing flexibility in QD formation. This axial confinement can be introduced in several ways: local electrostatic gates \cite{Fasth2005, Pfund2006},
 heterostructures \cite{Cornia2019},
crystal-phase modulation \cite{Nilsson2016, Debbarma2022, Ungerer2024}, controlled doping profiles \cite{Yang2005, Lauhon2002}, etching techniques \cite{Fulop2016, Berg2016, Seidl2021}, or by contact-defined tunnel barriers at the metal–NW interfaces \cite{ Fan2015, Taupin2016}. 

Among the III–V semiconductors, indium arsenide (InAs) is a particularly well-studied platform for quantum transport \cite{Nilsson2016,  Cornia2019, Nilsson2016_2, Debbarma2022, Ungerer2024, NadjPerge2010, Schroer2011, Chang2015, Krogstrup2015}.
Its low effective electron mass and narrow band gap enhance quantization effects and spin–orbit coupling, and donor-like surface states facilitate the formation of ohmic contacts \cite{Olsson1996, Bjork2005, Csonka2008}.
Recent experiments have shown that the hexagonal cross-section of InAs NWs, together with the intrinsic surface accumulation layer, facilitates the formation of nearly rotationally symmetric QDs \cite{Potts2019, Debbarma2021, Debbarma2022, Aspegren2024}. 
In such systems, the effective Landé \(g^\ast\)-factor can be tuned from nearly zero to values exceeding 100, for Zeeman energies smaller than the spin–orbit coupling \cite{Winkler2017}.
Spin–orbit coupling energies approaching 1 meV have been reported in crystal‐phase engineered InAs NWs \cite{Aspegren2024}, potentially enhanced by polarization charges at the wurtzite–zinc blende (WZ-ZB) interfaces \cite{Chen2017,Belabbes2013,Hjort2014,Li2014,Bauer2014,Kriegner2011}. 
While a large spin–orbit coupling is attractive for electrically controlled spin qubits, it also demands strong confinement to avoid unwanted orbital mixing. 
An increased radial confinement can be achieved either by gate-induced depletion,
such as from a wrap-around gate \cite{Tarucha1996}, or simply by growing
thinner NWs. However, the latter is challenging using crystal-phase modulation, as it becomes increasingly difficult to form the ZB crystal phase in very thin InAs nanowires \cite{Dick2010}.

In this work, we explore strongly confined QDs realized by a hybrid approach of combining strong axial confinement via epitaxial axial barriers and reduced radial dimensions via isotropic etching.
This allows us to circumvent the physical limitations on QD formation through crystal-phase control.
We study how the charging energy, governed by capacitive contributions, scales with QD dimensions and compare experimental results with finite-element-method simulations.
The resultant three-dimensional confinement potential supports well-resolved orbital spectra and significantly enlarged charging energies for the same charge configurations.
Our method provides a flexible platform for investigating few-electron effects and spin dynamics in QDs with hard-wall confinement. 
Such features are particularly relevant for the development of spin–orbit qubits and other spin-based quantum technologies, where well-controlled quantum states are required.

\begin{figure}
    \centering
    \includegraphics{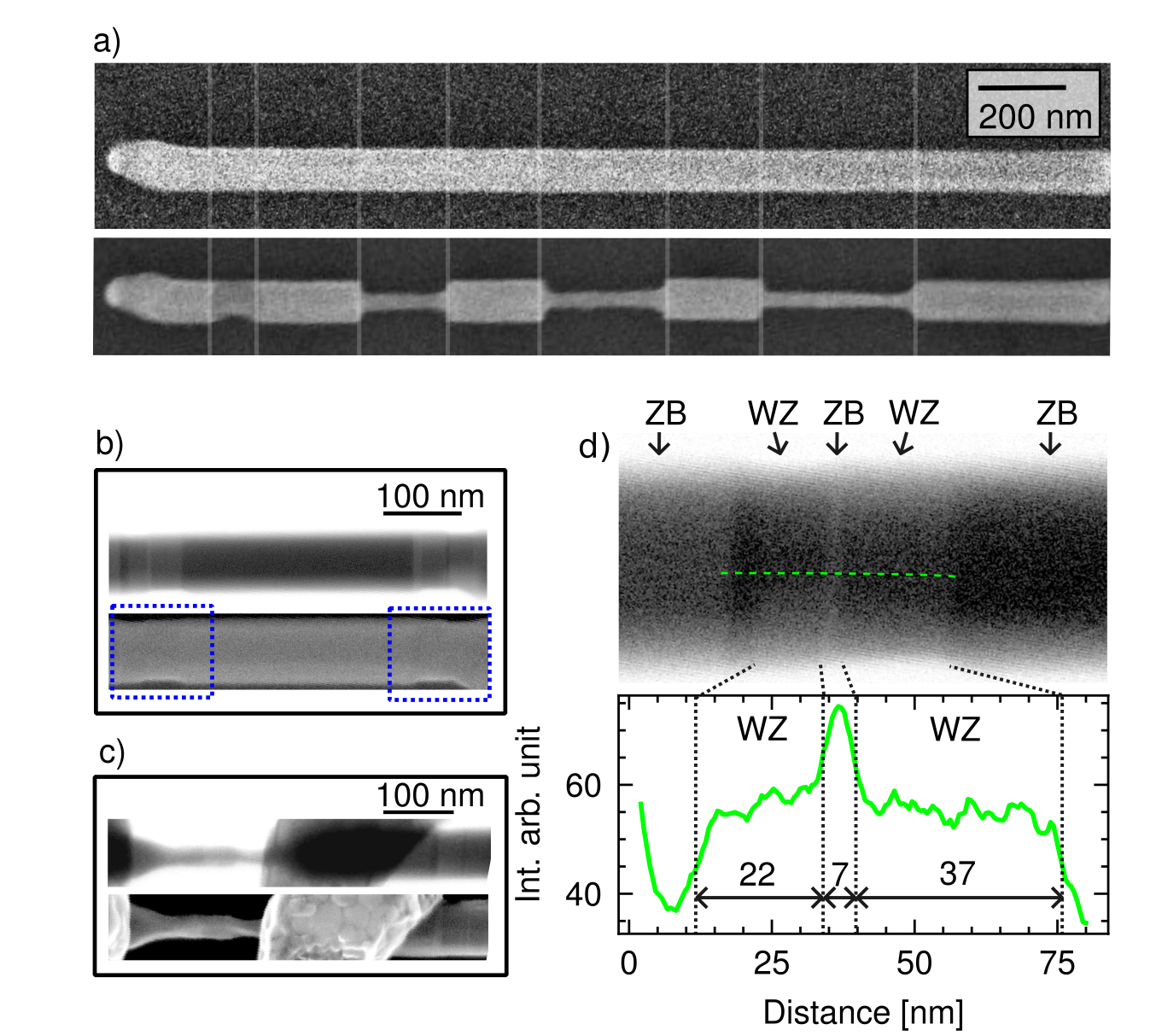}
    \caption{
        a) SEM images before and after etching in a diluted citric acid (C$_6$H$_8$O$_7$) and hydrogen peroxide (H$_2$O$_2$) mixture for 15 s. Etching procedure and mixture details are found in the main text.
    b) ECCI and SEM micrographs of a representative NW hosting two QDs, with crystal-phase dependent contrast revealing the zinc blende (ZB) and wurtzite (WZ) segments. 
    c) ECCI and SEM micrographs of an etched QD segment (left, corresponding to device D2) and an unetched QD (right). The ECCI methodology reveals clear contrast differences between the crystal phases for segments with well-defined crystal facets. However, the small interaction volume in etched segments prevented extraction of their dimensions in this work.
    d) Higher resolution ECCI micrograph of the left QD in b, with extracted dimensions. 
    ZB rotational twin planes reverse the contrast, which is observed in the left ZB segment.}
    \label{fig:1}
\end{figure}

\section{Main}
We base our experiments on InAs NWs with crystal-phase-defined barriers deposited on a SiO$_2$ substrate with an underlying global Si back-gate.
The hard-wall barriers that define the QD's confinement potential are determined during growth and consist of two WZ barriers sandwiching a thin ZB segment.  
Scanning electron microscopy (SEM) shows a spread in NW and QD dimensions, with diameters of 75-90 nm, dot lengths of 4-12 nm, and barrier lengths of 20-40 nm.
An example of such an extraction is shown in Figure \ref{fig:1}c, where electron-channeling contrast imaging (ECCI) shows varying contrast for the ZB and WZ phases due to crystal-phase-dependent backscattering \cite{Zaefferer2014, Nilsson2016}.
This example QD has barrier lengths of $22$ and $37$ nm and a dot length of $7$ nm.
The NWs used in this work were grown with two spatially separated QDs in the axial direction (see Figure \ref{fig:1}b-c). 
For some NWs, this was used to create an unetched reference QD in the same NW. 

Prior to etching, we locate the QD by identifying an indent in the NW, where the WZ barriers suppress radial overgrowth. 
This indent is visible in the NWs in Figures \ref{fig:1}b-d, but not for the NW presented in Figure \ref{fig:1}a.
For NWs with a visible indent, we lithographically define an opening of 180 nm centered on the indent, followed by the etching procedure.
Two etching times were used, 15 and 17 s.
The post-etch QDs had a spread of $d_\mathrm{NW}$ $22-42$ nm, as compared to the pre-etched NW diameters of $77-84$ nm.
Here, the pre-etched diameters refer to the diameter of the NW segments surrounding the QD, i.e., not the QD itself, while the post-etched diameter refers to the QD.
This discrepancy is a consequence of recording the pre-etch SEM images at low resolution to minimize the effects of carbon-based deposition on the etch rate.

The etchant solution was prepared by dissolving 1g of monohydrate citric acid (C$_6$H$_8$O$_7$) in 100mL of H$_2$O, which was subsequently mixed with H$_2$O$_2$ at a 5:1 ratio.
We found an average radial etch rate of 1.9$\pm$0.2 nm/s for InAs, about twice the reported value in \cite{Seidl2021}, which may be explained by geometric differences between the etched samples, as the multi-faceted NW morphology potentially leads to faster etching than in planar geometries.
NWs with pre-existing indents typically had a more uneven surface after the etch, as shown in Figure \ref{fig:1}c and in the Supplementary.
Here, we speculate that the additional facets around the ZB/WZ/ZB indents have different surface energies, leading to a non-uniform etch rate and further altering of the morphology.

\begin{figure*}[htbp]
    \centering
    \includegraphics{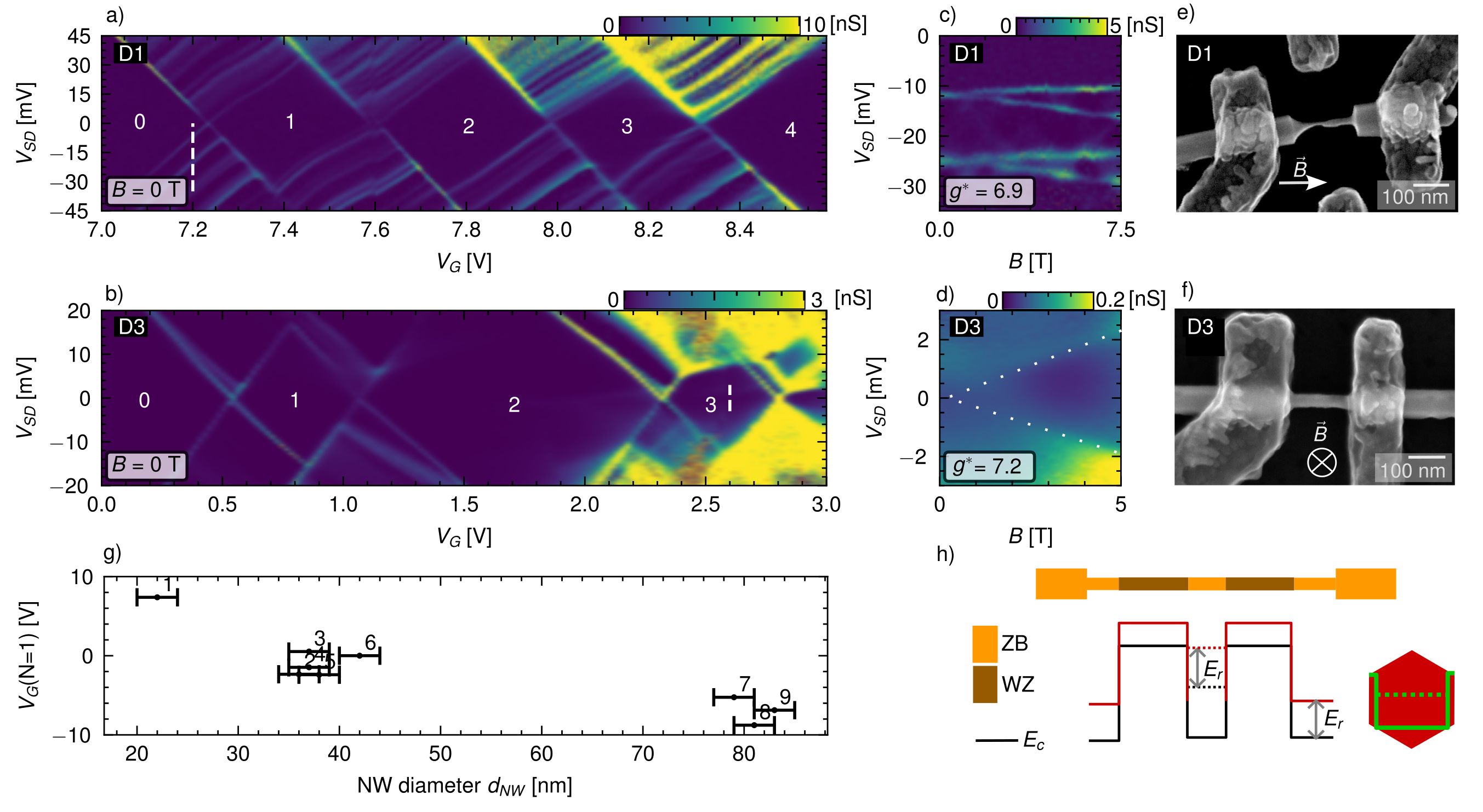}
    \caption{a, b) Stability diagrams of two etched QDs near depletion at $B=0$. 
    c,d) Magnetic field dependence of the 1e charge state, where the field is applied along the direction shown in e,f. The $V_{SD}$ sweep in c is measured along the dashed line in a, while d is along the dashed line in b. 
    e,f) SEM images of the two devices, where the thinned-down QDs are connected to electrodes. The sidegates visible in e are not used. 
    g) Gate voltage at which the first electron occupies the QD, i.e., $V_G(N=1)$ versus NW diameter. Confinement and screening effects result in a positive shift with a reduced cross-section.  
h) Schematic showing that the subbands in both the WZ and ZB phases gain energy from the radial confinement, delaying the loss of axial confinement for thin diameters. }
    \label{fig:experimental}
\end{figure*}

In Figure \ref{fig:experimental} we present measurements on two QDs with post-etch diameters of 22 nm and 37 nm; see Figure \ref{fig:experimental}e,f.
In  Figure \ref{fig:experimental}a,b charge stability diagrams show the electron filling of the two QDs, as indicated by the characteristic Coulomb diamonds.
Numbers between 0 and 4 label the number of electrons residing on the QD, denoted $N$e. 
We use the same methods to extract device parameters across all devices, focusing on charging energies, capacitive couplings, and gate lever arms for the 1e charge configuration.

The charging energy $E_C$ is set by the capacitive couplings between the QD and its surroundings, $E_C = e^2 / C_\Sigma$, where the dominant contributions tend to be the surrounding electrodes, i.e., $C_\Sigma \approx  C_S + C_D + C_G$, where $S, D, G$ denote the source/drain contacts and any gates present in the system. 
We find charging energies $E_C$ of 33 and 18 meV for devices D1 and D3, which are considerably higher than those of the unetched reference QDs (average $E_C<10$ meV).
The gate capacitance is extracted from the diamonds as $e/\Delta V_G$, where $\Delta V_G$ is the diamond width.

Directly related to the changes in $E_C$ and $C_G$, we find a corresponding difference in gate levers between the two devices, with $\alpha_G = 0.09$ and $0.03$, where we define the gate lever arm $\alpha_G = {C_G}/{C_\Sigma}$ (summarized for all devices in Figure \ref{fig:fig3}c).
These lever arm values suggest that the source-drain leads dominate the capacitive contributions, and we will show that the increase in both $E_C$ and $\alpha_G$ is, as expected, primarily driven by a reduction in $C_{S,D}$ as the QD diameter decreases.

Outside the Coulomb diamonds, sharp resonances arising from excited states in both the QD and leads are seen. 
As the semiconductor leads are etched down along with the QD, the density of lead states per unit energy decreases. 
Consequently, the broadening effect of such features, typically seen in thicker NWs due to many overlapping lead states, is suppressed.

\begin{table*}[t]
\centering
\begin{tabular*}{\textwidth}{@{\extracolsep{\fill}} l c c c c c c c}
\hline
    ID & $d_{NW}$ [nm] & $V_{G}(N=1)$ [V] & $E_C$ [meV] & $\Delta E_{1-0}$ [meV] & $C_G$ [aF] & $C_{S+D}$ [aF] & $g^*$ \\
\hline
D1 & $22 \pm 3$ & 7.4  & $33  \pm 2$ & $5 \pm 2$   & $0.43 \pm 0.04$ & $4.42 \pm 0.36$ & 6.9$_\parallel\pm 0.27$ \\
D2 & $36 \pm 3$ & -2.3 & $21  \pm 2$ & $21 \pm 2$  & $0.64 \pm 0.08$ & $7.00 \pm 0.90$ & 4.8$_\perp\pm 0.25$ \\
D3 & $37 \pm 1$ & 0.5  & $18  \pm 2$ & $16 \pm 2$  & $0.30 \pm 0.02$ & $8.86 \pm 1.30$ & 7.2$_\perp\pm0.27$ \\
D4 & $37 \pm 1$ & -1.8 & $17  \pm 2$ & $16 \pm 2$  & $0.32 \pm 0.02$ & $9.10 \pm 1.38$ & 5.8$_\perp\pm 0.27$ \\
D5 & $38 \pm 5$ & -2.4 & $22  \pm 2$ & $18 \pm 2$  & $0.51 \pm 0.05$ & $6.78 \pm 0.82$ & 8.6$_\perp\pm 0.26$ \\
D6 & $42 \pm 1$ & 0.0  & $13  \pm 1$ & $12 \pm 1$  & $0.34 \pm 0.01$ & $12.0 \pm 1.18$ & 7.7$_\perp\pm 0.34$ \\
\hline
D7 & $79 \pm 1$ & -5.3 & $11 \pm 2$ & $12 \pm 2$ & $0.40 \pm 0.02$ & $14.2 \pm 1.32$ &  -\\
D8 & $81 \pm 1$ & -8.2 & $10 \pm 1$ & $7 \pm 1$  & $0.49 \pm 0.02$ & $15.5 \pm 1.63$ &  -\\
D9 & $83 \pm 1$ & -6.9 & $8 \pm 1$  & $5 \pm 1$ & $0.56 \pm 0.03$ & $19.5 \pm 2.48$ & -\\
\hline
\end{tabular*}
\caption{Summary of extracted parameters for nine devices used in this study. Samples labeled D1 and D3 are presented in Figure \ref{fig:experimental}. Error margins are estimated as follows: $d_{NW}$ limited by SEM resolution and $d_{NW}$ variation over QD segment, $E_C$ and $\Delta E_{1-0}$ from linewidth broadening of spectral lines, $C_G$ propagated from accuracy in $\Delta V_G$ extraction, and $C_{S+D}$ propagated from error in $E_C$ and $C_G$.
    Devices D7–D9 correspond to unetched devices. The $\perp/\parallel$ notation refers to field orientation, $\perp$ is applied perpendicular to the NW axis, and $\parallel$ approximately parallel to the NW axis.}
\label{table:1}
\end{table*}

We extract $g$-factors (see Figure \ref{fig:experimental}c,d) for some electron fillings in order to see whether the values show indications of quenching due to the strong confinement \cite{Bjork2005}.
When possible, the $g$-factor was extracted inside a Coulomb diamond using second order tunneling (Figure \ref{fig:experimental}d), otherwise using first order tunneling (Figure \ref{fig:experimental}c) where we also compensate for tunnel barrier asymmetry.
The range of extracted $g$-factors is between $5$ and $9$, with no clear dependency on $d_{NW}$.
We note that the extracted $g$-factors should not be strongly affected by orbital contributions as either the s-like orbital is probed, or the $B$-field is perpendicular to the NW axis. 
The values vary around typical $g$-factors observed for InAs QDs reported by many groups \cite{Bjork2005, Pfund2006, Hansen2005, Fasth2007, Nilsson2016, Potts2019}.
We therefore conclude that there is no strong $g$-factor suppression of the QDs in this study, which indicates room for further downscaling.
A summary of the extracted parameters for nine devices is shown in Table \ref{table:1}, where we also introduce an orbital excitation energy $\Delta E_{1-0}$, which is extracted as the height difference between the 1e and 2e diamonds, and $V_{G,1}$ which is the gate voltage where the first electron occupies the QD.

In the transport measurements, we generally observe a substantial increase in the tunneling current at the onset of transport through a new orbital, especially for the 3e charge state. The combination of shallow axial barriers and a large orbital energy separation results in a significant change in tunneling probability. 
Alternatively, a sharp rise can also be associated with the onset of transport involving higher-index axial levels, which have an increased barrier overlap and higher probability for tunneling \cite{SadreMomtaz2020}. 
Despite the significant difference in radial and axial QD dimensions, the energy difference between radial and axial excitations can be comparable due to differing barrier energies.
This could explain the lack of a clear trend in $\Delta E_{1-0}(d_{NW})$ (displayed in Table \ref{table:1}) and a reduced quenching of $g$-factors. 

\section{Simulation}

For disk-shaped QDs in NWs,  \( C_S + C_D \) typically dominates over \( C_G \) by approximately one to two orders of magnitude depending on gate geometry and QD dimensions.
However, as the QD radial dimension is reduced, contributions beyond the cross-sectional area of the NW become increasingly significant.
From purely semiconductor and geometrical perspectives, $C_G$ is expected to grow approximately linearly with the axial QD length, and change as the charge centroid is shifted with gating.
Meanwhile, the capacitive coupling to the NW leads, $C_{S,D}$ could be expected to be proportional to the cross-sectional area.
However, $C_{S,D}$ incorporates contributions from stray capacitances from metallic contacts and the semiconductor leads, where the latter depend on the charge distribution in the NW leads.

\begin{figure*}[htbp]
    \centering
    \includegraphics{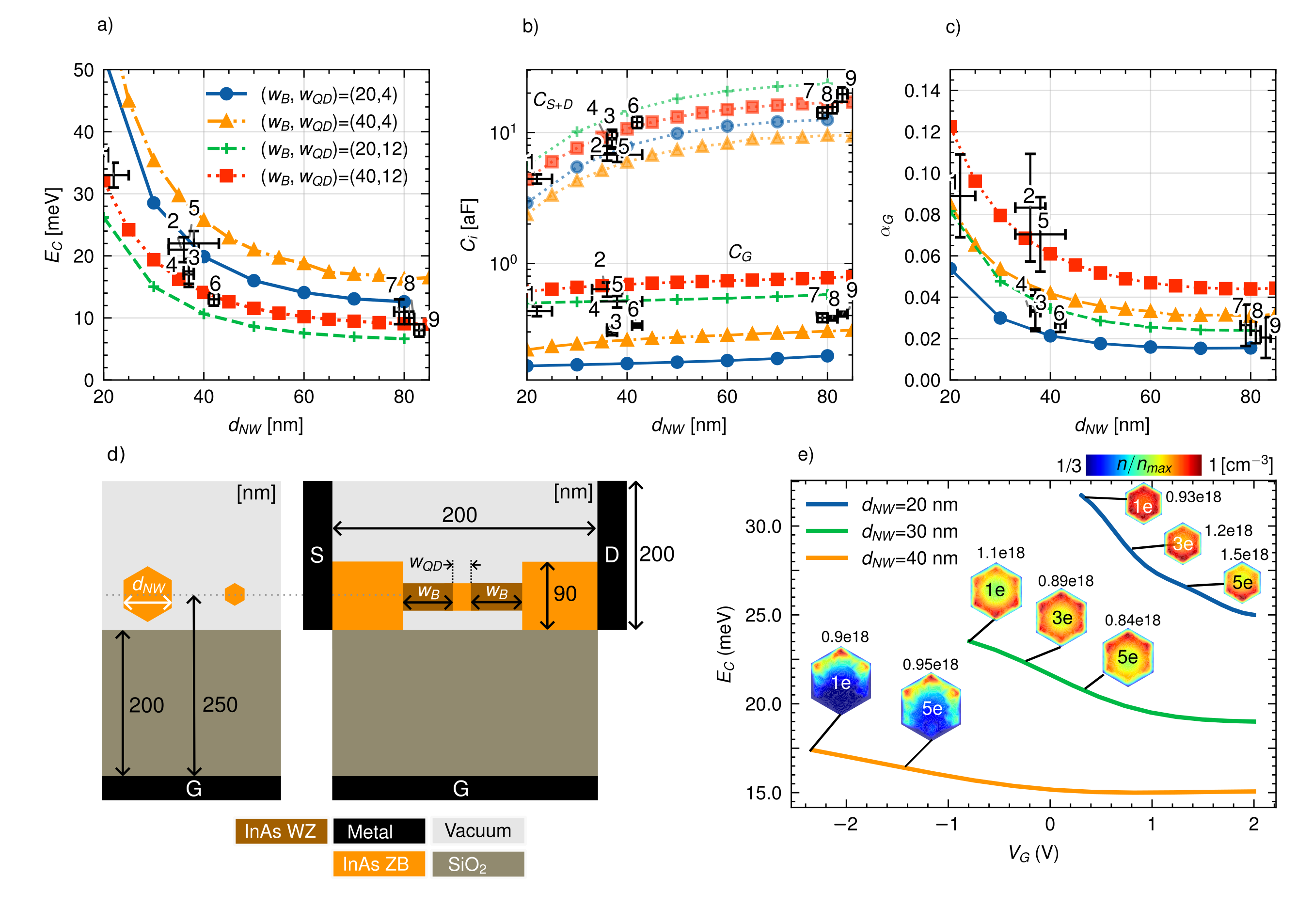}
    \caption{
    a-c) Simulated charging energies, capacitances, and gate lever arms as a function of radial extent of QDs.
    Legend in a refers to QD dimensions.
    Solid markers correspond to experimental results, where numbering corresponds to the devices in Table \ref{table:1}.
    d) Schematics showing parameters for the simulation.
    A NW with varying diameter, barrier, and dot length is suspended a distance away from a back-gate.
    The barriers are treated as dielectrics, and the capacitances are extracted for each electrode (black).
    e) Charging energy as a function of gate bias for a 20 nm (blue), a 30 nm (green), and a 40 nm (orange) thick QD, with $w_B=40$ nm and $w_{QD}=12$ nm. The color maps show the electron distribution at certain charge configurations. Each colormap is normalized to itself, with the maximum value $n_{max}$ included next to each plot.}
   \label{fig:fig3}
\end{figure*}

Near depletion, especially in large diameter QDs, the charge centroid shifts away from the back-gate, leaving a depleted semiconductor between the gate and the confined charge.
The smaller effective QD diameter results in a lower $C_{ SD }$. 
However, the effect is to a large degree offset by screening due to the significant dielectric constant of the depleted InAs. 
As shown in Figure \ref{fig:fig3}a, such depleted segments significantly reduces Ec (N=1) in large $d_{ NW }$ QDs as compared to etched QDs.

The geometric and semiconductor properties are accounted for by solving a finite-element model in COMSOL, where quantum confinement effects are included via the density-gradient method \cite{Ancona2011}.
We use a 3D model of a NW placed on top of a gate dielectric.
Two metallic plates ($200\cross200$ nm) at the boundaries of the simulated room are connected to two InAs ZB segments, each with a fixed diameter of $90$ nm.
Suspended in between the two ZB segments is another, thinner, NW that is further divided into three segments consisting of a ZB disk sandwiched between two WZ barriers. 
The two WZ barriers are treated as dielectrics.
In the simulations, symmetric barriers are assumed, and the barrier width ($w_B$), NW diameter ($d_\mathrm{NW}$), and QD length ($w_{QD}$) are varied around experimental values, as depicted in Figure \ref{fig:fig3}d.
The ZB to WZ barrier height is set to 0.1 eV, and 4.9 eV for the QD-to-vacuum interface.
The gate dielectric is a 200 nm thick SiO$_2$ layer, separating the NW from a global backgate.
The total length of the junction along the NW axis is 200 nm. 
As spatial parameters are changed, the NW axis is kept at a fixed distance from the gate contact and dielectric, consistent with the experimental geometry.
The model captures the gating influence on $E_C$ by reshaping the charge cloud in the QD and leads, thereby changing dimensions relevant for capacitive coupling(s), but fails to capture, e.g., the effective change in $w_{QD}$ resulting from altered barrier heights.
A complete treatment of these effects would require a full self-consistent Schroedinger-Poisson model.

The surface charge accumulation, and the resulting tubular charge distribution, was found to be particularly significant for matching simulations and the experimental data.
The best match was found for a surface donor density of $5\cdot 10^{12}$ cm$^{-2}$, along with a weak background donor doping of $N_d = 10^{16}$ cm$^{-3}$.
These values match reference values for InAs systems, where a Fermi level pinning between 0.06 and 0.16 eV above the conduction band edge results in sheet charge densities of $10^{12}-10^{13}$ cm$^{-2}$ \cite{Olsson1996, Degtyarev2017}.

The total charge on the QD for a given bias $Q(V)$ is found by integrating the space charge density 
$$\rho(V) = q \cdot [N_d - n(V) + p(V)]$$ 
over the QD region, where $n,p$ correspond to electron and hole contributions.
For Figures \ref{fig:fig3}a-c, we extract the capacitances around the gate bias for which the total integrated charge approaches \(Q(V_G) = -e\) by applying a small bias $\Delta V_{i}$ on top of the depleting $V_G$ bias via
$$C_i = \frac{\Delta Q}{\Delta V_i} $$
where $\Delta Q$ is the change in charge residing on the dot as induced by the small bias perturbation $\Delta V_i$ on the source, drain, or gate. 
This approach, similar to what was used in \cite{Khanal2007}, enables us to investigate how each capacitive coupling \(C_{i}\) varies with $V_G$ and QD dimensions.  

In Figure \ref{fig:fig3}a experimentally found $E_C(N=1)$ are overlayed with simulated systems.
Four combinations of dot and barrier lengths are included in the simulation, corresponding to the range of dimensions extracted via electron microscopy for the NW growth. 
Here, we find excellent agreement with the experiments, where the charging energy increases as the QD is radially thinned down.
The trend in $E_C(N=1)(d_{NW})$ deviates from $E_C \propto 1/d_{NW}^2$, mainly due to stray capacitances and depletion effects in the semiconductor leads.
Thin QDs deplete at nominally negative or positive gate voltages. Hence, at the threshold voltage, the corresponding depletion in the NW leads is negligible, resulting in a system close to the geometric maximum.
As such, the expected trend for narrow ($<45$ nm) QDs would be $E_C \propto 1/d_{NW}^2$. However, we find that the trend is partially quenched by stray capacitances present in the system, e.g., those arising from unetched NW segments and metallic leads, which are included in the simulations.

In contrast, QDs with a large $d_{NW}$ deplete at high negative voltages, thereby reducing the carrier concentration in the semiconductor leads and their contribution to $C_{S,D}$.
Geometric ($E_C \propto 1/d_{NW}^2$) and depletion effects compete, resulting in a saturation in $E_C(N=1)$ for $d_{NW}>45$ nm.
Hence, having a long etched segment and thin metallic contacts placed far away is necessary to maximize $E_C$.
In the Supplementary we show corresponding simulations where contributions from stray capacitances are minimized.

The voltage at which the first electron occupies the dot, $V_{G}(N=1)$, largely depends on the degree of confinement.
In addition, $E_C$ is typically lowered for subsequent charge states.
This is studied in Figure \ref{fig:fig3}e, where the charging energies and electron distribution for charge states near depletion are shown for three different QD diameters, with $w_B=40$ and $w_{QD}=12$ nm.
Here, we find that depletion occurs at increasingly positive gate voltages as the NWs become thinner, a consequence of (radial) quantum confinement pushing subbands to higher energies.
This trend agrees qualitatively with the diameter dependence of $V_{G}(N=1)$ in Figure \ref{fig:experimental}g, where the absolute values differ but follow the same overall dependence on $d_{NW}$.

Small systems exhibit enhanced sensitivity to surface effects due to their large surface-to-volume ratio. 
This effect is particularly pronounced in InAs-based systems, where the downward band bending of the conduction band at the surface pushes the electronic wavefunction towards the device edges. 
Since the surface is a significant source of scattering, our strategy of physically shrinking the QDs may lead to poor candidates for quantum information applications.
This limitation could possibly be mitigated by improved etching techniques, such as digital etching \cite{Berg2016}, combined with post-etch surface passivation.
An alternative approach could be in-situ dry etching during or immediately after growth, as previously demonstrated for tuning of NW radial dimensions in vapor-phase epitaxy and molecular beam epitaxy \cite{Borgstrom2010, Loitsch2015}.
Applying such an approach in combination with crystal-phase-defined QDs, followed by a heating cycle and the growth of a radial passivating shell, could enable strong confinement while simultaneously reducing surface scattering. 

\section{Conclusion and outlook}
We demonstrated a hybrid approach to fabricate QDs embedded in NWs by combining epitaxial crystal-phase engineering and chemical wet etching. 
This methodology resulted in QDs with a considerably reduced cross-section and with strong confinement in all directions.

We found that both charging energies and gate lever arms increased as the cross-section was reduced, which is useful for tuning QD systems with well-defined charge occupation at higher temperatures.
A reduced cross-section also results in QDs that require less extreme voltages to deplete, simplifying device operation.
In addition, minimizing the electrical-field contributions from a gate facilitates quantum transport studies of Rashba-induced spin-orbit interaction at WZ/ZB interfaces.

By performing finite-element simulations, we highlighted limitations in gate-defined approaches for enhancing $E_C$, and found a regime where stray capacitances limit further $E_C$ enhancement from cross-section reduction alone. 
In all, these findings establish a framework for enhancing confinement in hard-walled NW QDs, which is relevant for advancing spin-orbit qubit technologies.

\vspace{2em}
\vspace{2em}

\section{Acknowledgements}
 This work was supported by the Knut and Alice Wallenberg Foundation, the Göran Gustafsson Foundation, the Swedish Research Council (VR 2024-04334, 2023-04142), The Royal Physiographic Society in Lund, and NanoLund.

\section{Data availability statement}
The data supporting the findings of this study are available upon reasonable request from the authors.
 
\bibliographystyle{unsrtnat}
\bibliography{references}

\pagebreak
\appendix
\renewcommand{\thesection}{S.\arabic{section}}
\renewcommand{\thefigure}{S\arabic{figure}}
\renewcommand{\thetable}{S\arabic{table}}
\setcounter{figure}{0}
\setcounter{table}{0}

\section{Supplementary Information}

\subsection{Etching procedure}
We prepare the etchant by first dissolving \textbf{1 g} citric acid monohydrate (C$_6$H$_8$O$_7\!\cdot$H$_2$O) in 100~mL H$_2$O to make a stock solution, and then mixing this stock with H$_2$O$_2$ in a \textbf{5:1} (v/v) ratio. 
Initial etch tests indicated an etch rate of approximately 1.9~nm/s (Table~\ref{tab:etchrate}). 
Variations in etched QDs correlate with local morphology: near-crystal-phase indents (WZ/ZB/WZ), additional facets can appear and promote locally faster etching, thereby modifying the post-etch morphology (see Fig.~\ref{fig:sup_sem}).

\begin{table*}[b]
\centering
\caption{Measured etch rates for InAs NWs in citric acid/H$_2$O$_2$ solution. This data does not correspond to devices shown in this paper. The etch rate (per side) is calculated from the diameter reduction divided by the etch time. Uncertainty corresponds to the sample standard deviation across eight devices.}
\label{tab:etchrate}
\begin{tabular}{c c c c c c}
\hline
\textbf{Before (nm)} & \textbf{After (nm)} & \textbf{Etched (nm)} & \textbf{Etched/side (nm)} & \textbf{Etch Time (s)} & \textbf{Etch Rate (nm/s)} \\
\hline
77 & 31 & 46 & 23  & 12 & 1.9 \\
72 & 27 & 44 & 22  & 12 & 1.9 \\
76 & 31 & 45 & 23  & 12 & 1.9 \\
79 & 31 & 48 & 24  & 12 & 2.0 \\
72 & 26 & 46 & 23  & 12 & 1.9 \\
91 & 41 & 50 & 25  & 12 & 2.1 \\
88 & 39 & 49 & 24  & 12 & 2.0 \\
88 & 44 & 44 & 22  & 12 & 1.8 \\
\hline
\textbf{Mean} &  &  &  &  & \textbf{1.94} \\
\textbf{Std.\,dev.} &  &  &  &  & \textbf{0.09} \\
\hline
\end{tabular}
\end{table*}

\subsection{SEM}
\begin{figure*}[t]
    \centering
    \includegraphics[width=\textwidth]{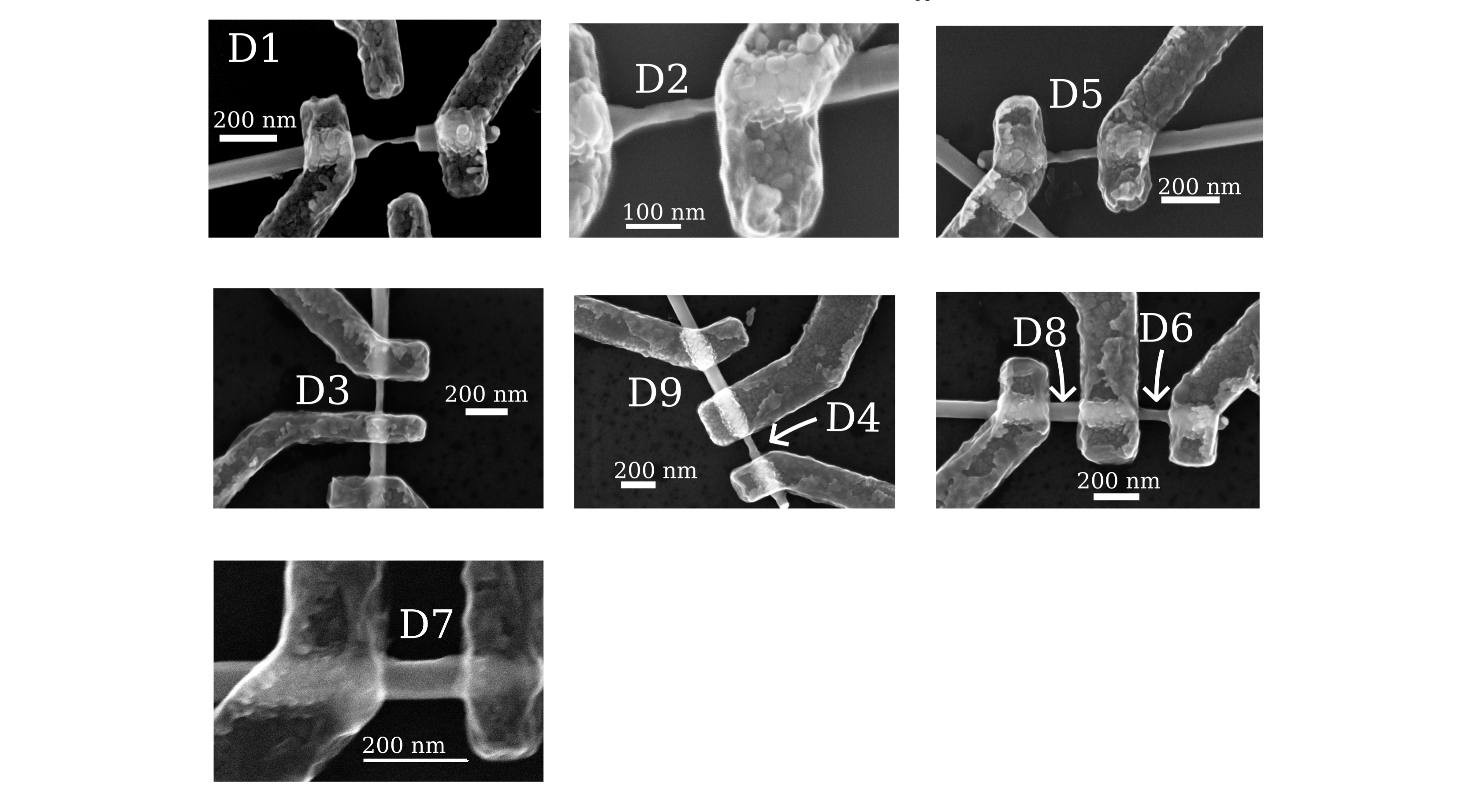}
    \caption{SEM images used for extraction of QD dimensions.}
    \label{fig:sup_sem}
\end{figure*}

SEM images of all devices used in this work are shown in Fig.\ref{fig:sup_sem}. Several devices are configured as three-terminal junctions hosting an etched QD in one arm and an unetched reference QD in the other.

\subsection{Charge stability diagrams}
\begin{figure*}[t]
    \centering
    \includegraphics[width=\textwidth]{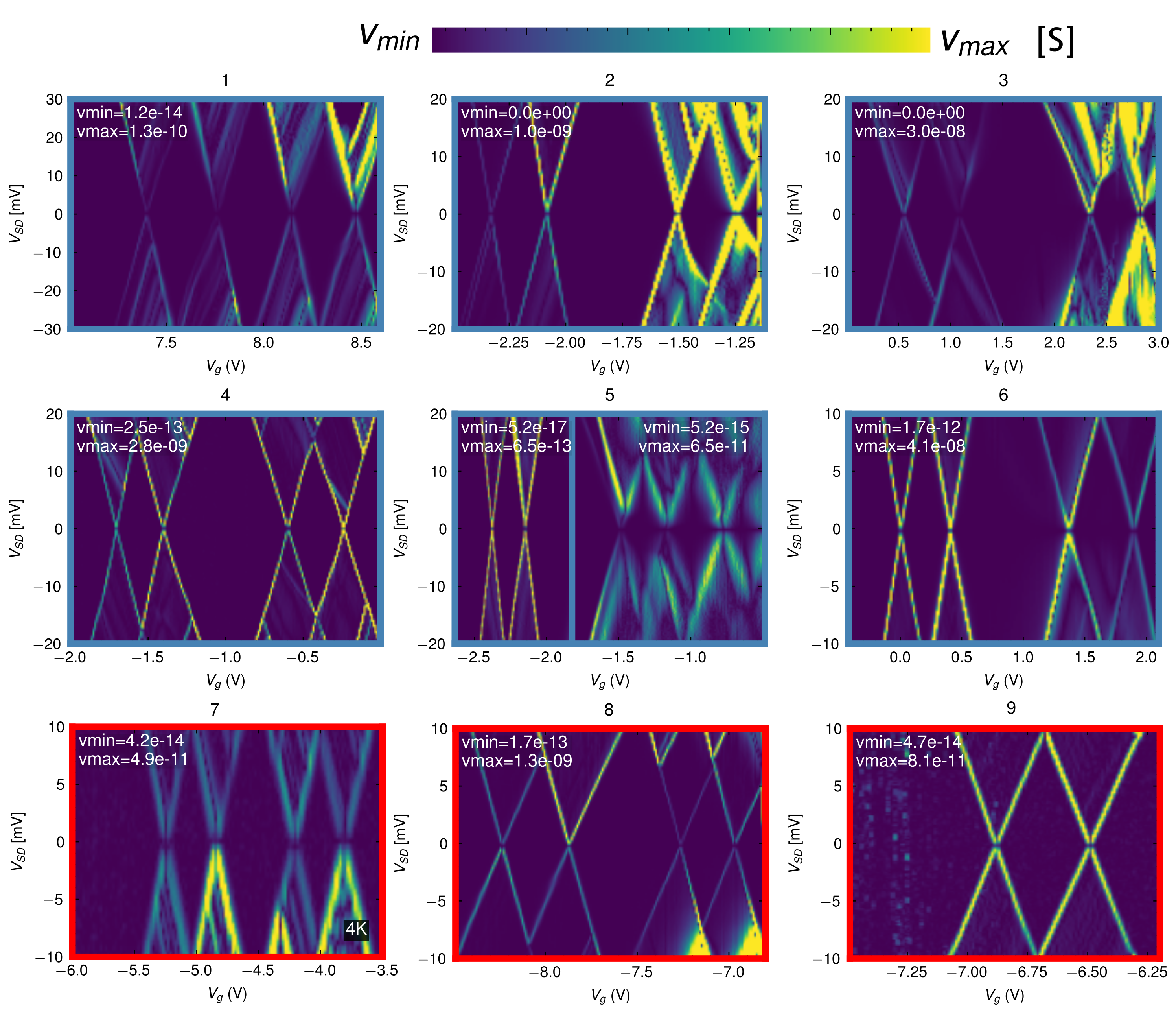}
    \caption{Charge stability diagrams of devices used to extract parameters in this work. Units in S.}
    \label{fig:sup_stab}
\end{figure*}

Charge stability diagrams for all devices used to extract parameters in this work are presented in Fig.~\ref{fig:sup_stab}. Each panel corresponds to a distinct device and shows characteristic Coulomb-blockade diamonds used to extract the charging energies ($E_C$), gate lever arms ($\alpha_G$), and capacitive couplings summarized in Table~I of the main text. Minor variations in the diamond slopes arise from device-specific tunnel-barrier asymmetries and local electrostatic environments.

\subsection{Minimized stray capacitances}
\begin{figure*}[t]
    \centering
    \includegraphics[width=\textwidth]{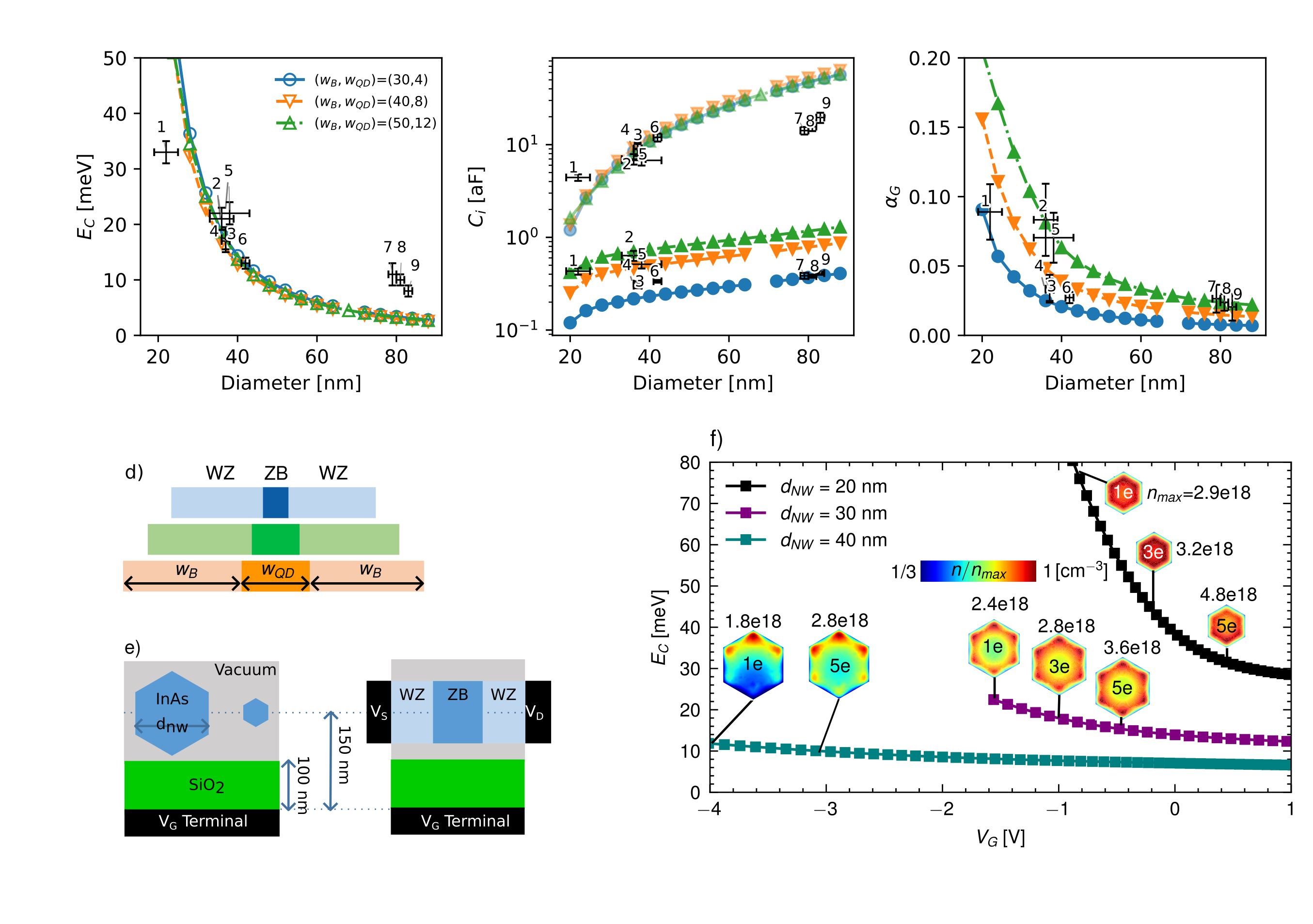}
    \caption{Simulations of $E_C$ and capacitive couplings for a system with minimized stray capacitances.}
    \label{fig:sup_model}
\end{figure*}

In the main text, we present simulations incorporating stray capacitances. By adjusting the model to minimize such contributions (Fig.~\ref{fig:sup_model}), we identify a much stronger $E_C \propto 1/d^2$ dependence, where $E_C$ in thin NWs increases significantly faster than when stray capacitances are included. Likewise, in this model, there is no depletion of the semiconductor leads, and the saturation in $E_C$ for large diameters is absent.

As in the main text, this configuration assumes WZ barriers as dielectrics, whereas the metallic contacts have the same cross-section as in the NW system and are directly contacted to the WZ segments.
Apart from these adjustments, all parameters are the same as in the main text.

\end{document}